\documentclass[10pt, conference]{IEEEtran}
\usepackage{nopageno}
\usepackage{calc}
\usepackage{multirow}
\usepackage{filecontents}
\usepackage{scrextend}
\usepackage{graphicx}
\usepackage{graphics}
\usepackage{pdfpages}
\usepackage{enumitem}
\usepackage{tikz}
\usepackage[hyphens]{url}
\usepackage{multicol} 
\usepackage{xcolor}
\usepackage[ruled,vlined]{algorithm2e}
\SetAlCapNameFnt{\footnotesize}
\SetAlCapFnt{\footnotesize}
\usepackage{mathtools}
\usepackage{algpseudocode}
\usepackage{amssymb}

\usepackage{amsmath}
\usepackage{verbatim}
\usepackage{footnote}
\makesavenoteenv{tabular}
\makesavenoteenv{table}
\usepackage{tabularx,ragged2e,booktabs}
\usepackage{array}
\usepackage{caption}
\usepackage{lscape}
\usepackage[utf8]{inputenc}
\usepackage{fancyhdr}

\usepackage{setspace}
\usepackage[normalem]{ulem}
\usepackage[font=small]{caption}
\usepackage{color, colortbl}
\definecolor{cadmiumgreen}{rgb}{0.0, 0.42, 0.24}
\newcommand\redsout{\bgroup\markoverwith{\textcolor{red}{\rule[0.5ex]{2pt}{0.4pt}}}\ULon}
\definecolor{orange}{rgb}{1,0.5,0}
\definecolor{darkgreen}{rgb}{0,0.5,0}

\newcounter{example}[section]

\newcounter{case}[section]

\newcounter{step}[section]

\makeatletter
\IEEEtriggercmd{\reset@font\normalfont\scriptsize}
\makeatother
\IEEEtriggeratref{1}

\begin{document}

\bstctlcite{IEEEexample:BSTcontrol}
\title{DynUnlock: Unlocking Scan Chains\\ Obfuscated using Dynamic Keys}

\author{
	{\large Nimisha Limaye$^{\dagger}$  and Ozgur Sinanoglu$^{\ddagger}$}\\
  {$^{\dagger}$\,Tandon School of Engineering, New York University $^{\ddagger}$\,New York University Abu Dhabi, UAE}\\
  \normalsize{\{nsl278, ozgursin\}@nyu.edu}
\vspace{-0.28in}
}

\maketitle
\renewcommand{\headrulewidth}{0.0pt}
\thispagestyle{fancy}
\pagestyle{fancy}

\cfoot{
\copyright~2020 IEEE.
This is the author's version of the work. It is posted here for your personal use.
	Not for redistribution.\\
	The definitive Version of Record is published in
	Proc. Design, Automation and Test in Europe Conference (DATE), 2020.
}

\begin{abstract}
Outsourcing in semiconductor industry opened up venues for faster and cost-effective chip manufacturing. However, this also introduced untrusted entities with malicious intent, to steal intellectual property (IP), overproduce the circuits, insert hardware Trojans, or counterfeit the chips. 
Recently, a defense is proposed to obfuscate the scan access based on a dynamic key that is initially generated from a secret key but changes in every clock cycle. This defense can be considered as the most rigorous defense among all the scan locking techniques. In this paper, we propose an attack that remodels this defense into one that can be broken by the SAT attack, while we also note that our attack can be adjusted to break other less rigorous (key that is updated less frequently) scan locking techniques as well. 
\end{abstract}

\section{Introduction}\label{sec:intro}
Many threats have risen as a result of outsourcing in the IC supply chain. Untrusted entities such as foundry and end-user can use tools at their disposal, to steal intellectual property (IP) by reverse engineering, insert hardware Trojans, counterfeit or overproduce the chip~\cite{big_hack,counterfeit, RE}. To thwart these attacks, logic locking is considered a viable design-for-trust solution. Earlier logic locking schemes were implemented with high output corruptibility in mind; with an incorrect key the chip should produce most erroneous results for the given input patterns~\cite{RLL,FLL}. However, this property was exploited by boolean satisfiability based attack known as the \emph{SAT attack} in retrieving the secret key of logic locking~\cite{SAT}. 

SAT attack identifies a distinguishing input pattern (DIP) which prunes the key search space; the attack continues till no such DIP is found and terminates with the correct key. High output corruptibility assists this attack as a larger portion of the key search space is pruned with each DIP. To thwart this attack scan obfuscation/encryption approaches~\cite{stream_vs_block, da2018preventing} were proposed which obfuscates the scan-in and scan-out data based on a key. As all the successful attacks, such as (App)SAT~\cite{AppSAT,SAT}, in logic locking literature relies on scan access, these defenses can be said to thwart logic locking attacks. However, they incur significant area and delay overheads, and mainly find applications in Crypto chips. To protect against IP piracy attacks, more general and scalable solutions are required.

\begin{figure*}
    \centering
    \includegraphics[width=1.0\textwidth]{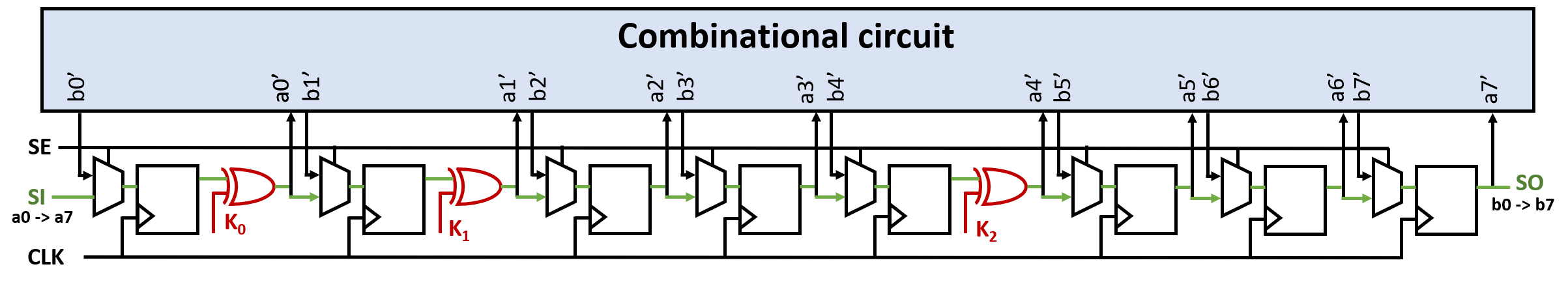}
    \caption{Scan obfuscation of $s208$ ISCAS-89 benchmark locked using three key bits.}
    \label{fig:eg}
\end{figure*}

Recently, XOR/MUX based scan obfuscation techniques have been proposed~\cite{EFF,DFS,DOS,ImprovedEFF} to obfuscate the scan data via XOR/MUX operations based on a secret key through key gates inserted between the scan flops. 
In~\cite{EFF}, a static key is used to obfuscate the scan chain content during scan in and out of patterns, effectively hindering unauthorized scan access. However, this defense was also recently broken by ScanSAT attack~\cite{scansat}.
In~\cite{DFS}, scan-out ports are blocked whenever the chip is in functional mode or upon a mode switch between functional and test mode, thereby, thwarting SAT-based attacks. However, this defense was also broken recently by shift-and-leak attack~\cite{shift-and-leak}.
Further, in~\cite{DOS,ImprovedEFF}, a dynamic scan locking key is used to obfuscate the scan data; the keys are generated using a Linear Feedback Shift Register (LFSR) based on a secret seed. 
In~\cite{DOS}, the key is updated upon scanning in and out a constant number ($p$) of patterns, while in~\cite{ImprovedEFF}, the key is updated in every clock cycle. The defense in~\cite{ImprovedEFF} can be viewed as the most rigorous, and thus, the most secure, dynamic scan locking defense. Indeed, the defense in~\cite{DOS} was broken recently even for its most rigorous version where the key is updated for every pattern ($p$=1)~\cite{ScanSAT_TETC}. Yet, the defense in~\cite{ImprovedEFF} remains unbroken.

\begin{table}[!tb]
\centering
\setlength{\tabcolsep}{2pt}
\small\addtolength{\tabcolsep}{5pt}
\footnotesize
\caption{Evolution of scan locking over last two years.}
\label{tab:compare}
\begin{tabular}{|l|c|c|c|}
\hline
    \textbf{Month, Year}   & \textbf{Defense} & \textbf{\begin{tabular}[c]{@{}c@{}}Obfuscation\\ type\end{tabular}} & \textbf{Attack} \\ \hline \hline
Jan, 2018   & EFF~\cite{EFF}              & Static                    & ScanSAT~\cite{scansat}         \\ \hline
May, 2018       & DFS~\cite{DFS}              & Static                    & Shift-and-leak~\cite{shift-and-leak}  \\ \hline
Sept, 2018 & DOS~\cite{DOS}              & Dynamic                   & Imp. ScanSAT~\cite{ScanSAT_TETC}        \\ \hline
May, 2019       & EFF-Dyn~\cite{ImprovedEFF}             & Dynamic                   & \textbf{This work}       \\ \hline
\end{tabular}
\end{table}

The evolution of scan locking defenses and attacks, focusing on the most recent years, is shown in Table~\ref{tab:compare}, illustrating the attention this area has been attracting lately.

\textbf{Contributions:} In this paper, we propose {a state-of-the-art} attack which can break all dynamic scan locking defenses~\cite{DOS,ImprovedEFF}. We present our \emph{DynUnlock} attack on \emph{the most rigorous case of dynamic locking}~\cite{ImprovedEFF} where the scan data gets obfuscated with a new key in every clock cycle. 
The contributions of this paper are as follows:
\begin{itemize}
    \item {We propose DynUnlock that can break all scan locking defenses.}
    \item We demonstrate our attack on a small ISCAS-89 dynamically scan locked circuit in Section~\ref{sec:attack} and perform it on six ISCAS-89~\cite{iscas89} and four ITC-99~\cite{itc99} benchmarks for 10 different LFSR seeds each.
    \item We aim to recover the secret LFSR seed generating the dynamic keys. We recover the unique seed for $70\%$ of the circuits, and for the remaining $30\%$ we obtain at most $128$ seed candidates out of $2^{128}$, which can be further refined very quickly via brute-force to recover the secret seed.
    \item The execution times of attack are maximum for $s38584$, $s38417$, and $s35932$ benchmarks ($<$ 7 minutes) as they have the maximum number of scan flops post-synthesis.
\end{itemize} 
\section{Background}\label{sec:bg}
SAT attack was a turning point for logic locking research; researchers have now started exploring approaches to thwart this fatal attack. Based on the observation that the success of SAT and other oracle-guided attacks relies on scan access, one promising approach is to obfuscate the scan data as it is delivered in and out of the scan chains. 

The defense in~\cite{ImprovedEFF} obfuscated the scan data dynamically by updating the key in every clock cycle (even within a pattern), resulting in the ultimate dynamic scan locking technique. In this paper, we propose an attack which will, \textbf{break all versions of dynamic scan locking}, including the most rigorous one~\cite{ImprovedEFF}. We use~\cite{ImprovedEFF} as a case study to showcase our attack methodology while we note that it is guaranteed to be successful on all other dynamic scan locking {versions} that are less rigorous than~\cite{ImprovedEFF}, where the key updates less frequently\cite{DOS}.

\begin{figure}
    \centering
    \footnotesize
    \includegraphics[width=0.45\textwidth]{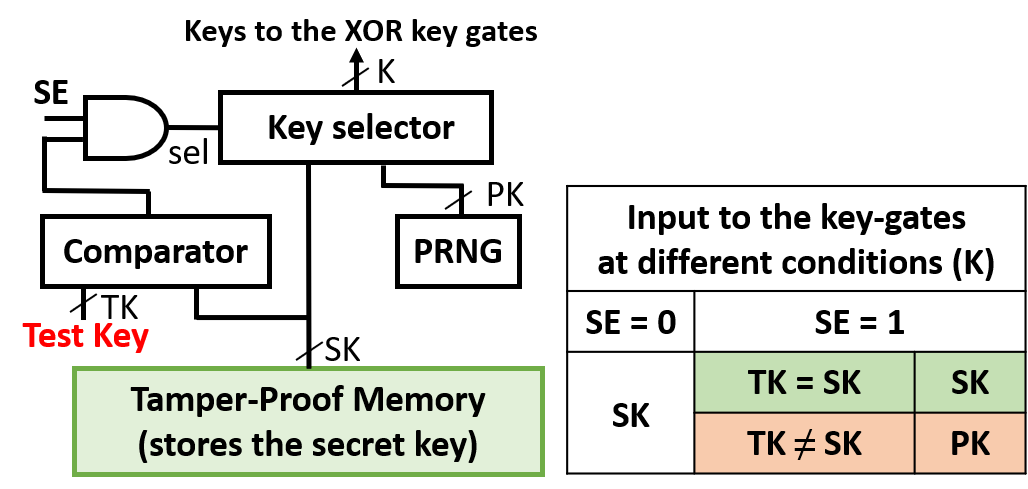}
    \caption{Test authentication scheme for EFF-Dyn. When test key (TK) mismatches with secret key (SK), comparator output goes low, and key selector passes the dynamic key from the pseudo-random number generator (PRNG) to the scan locked circuit. PRNG output updates every clock cycle.} 
    \label{fig:tas}
\end{figure}

\subsection{Case Study}
Dynamic Encrypt Flip-Flop (EFF-Dyn)~\cite{ImprovedEFF} combines {scan locking approach from~\cite{EFF} and a PRNG, to introduce dynamicity in the defense.}
During functional mode or the capture operation in test mode (scan enable ($SE$) signal is low), the secret {scan} locking key that is stored in the Tamper-Proof Memory (TPM) controls the key gates. During testing, an externally provided test key is expected. When this test key matches the secret scan locking key, the key gates receive this correct key during the scan shift operations ($SE$ signal is high) as well; in case of a mismatch, however, the PRNG that updates the key in every clock cycle controls the key gates dynamically. This is illustrated in Fig.~\ref{fig:tas}. 

With an incorrect scan locking key, the chip outputs will be {highly} corrupted. A testing facility, if trusted, can provide the scan locking secret key as the external test key to have this key (known to them) drive the key-gates during both shift and capture operations during test. Without the knowledge of this secret key, the access to the scan chains is locked based on a very dynamic scan locking key generated by the PRNG. 

\textbf{Security Properties.}
A dynamically obfuscated scan access is meant to prevent an attacker from applying the generated DIPs through the scan chains, as the PRNG would be updating the key in every clock cycle, {thereby, providing resilience against the SAT attack}. 
As a defense that {thwarts both SAT and AppSAT attacks,} EFF-Dyn seems to be a promising design IP protection technique.

\section{Proposed Attack}
% \label{sec:attack}
In this section, we discuss an attack approach on dynamically locked scan chains with EFF-Dyn as a case study. In our attack, \emph{DynUnlock}, we aim to recover the seed of the PRNG which produces the key sequence. With the secret seed known, an attacker can gain scan access without the knowledge of the scan locking secret key; an arbitrary test key can be used to leave the scan access control to the PRNG, which can be easily modeled by the attacker as long as its seed is known (see threat model below). 

\textbf{Threat Model.}
Consistent with the logic locking literature, we assume that the attacker %residing at a foundry 
can obtain the design netlist by reverse-engineering the GDSII files or a fabricated chip. This gives her access to all the structural information including the test authentication scheme, the location of key gates, and the PRNG structure and thus, its polynomial function.
We also assume that the attacker has access to a working oracle (functional IC). Though she also has access to scan ports physically, the attacker needs to get past the scan obfuscation defense.

\begin{figure}[!tb]
    \centering
    \includegraphics[width=0.4\textwidth]{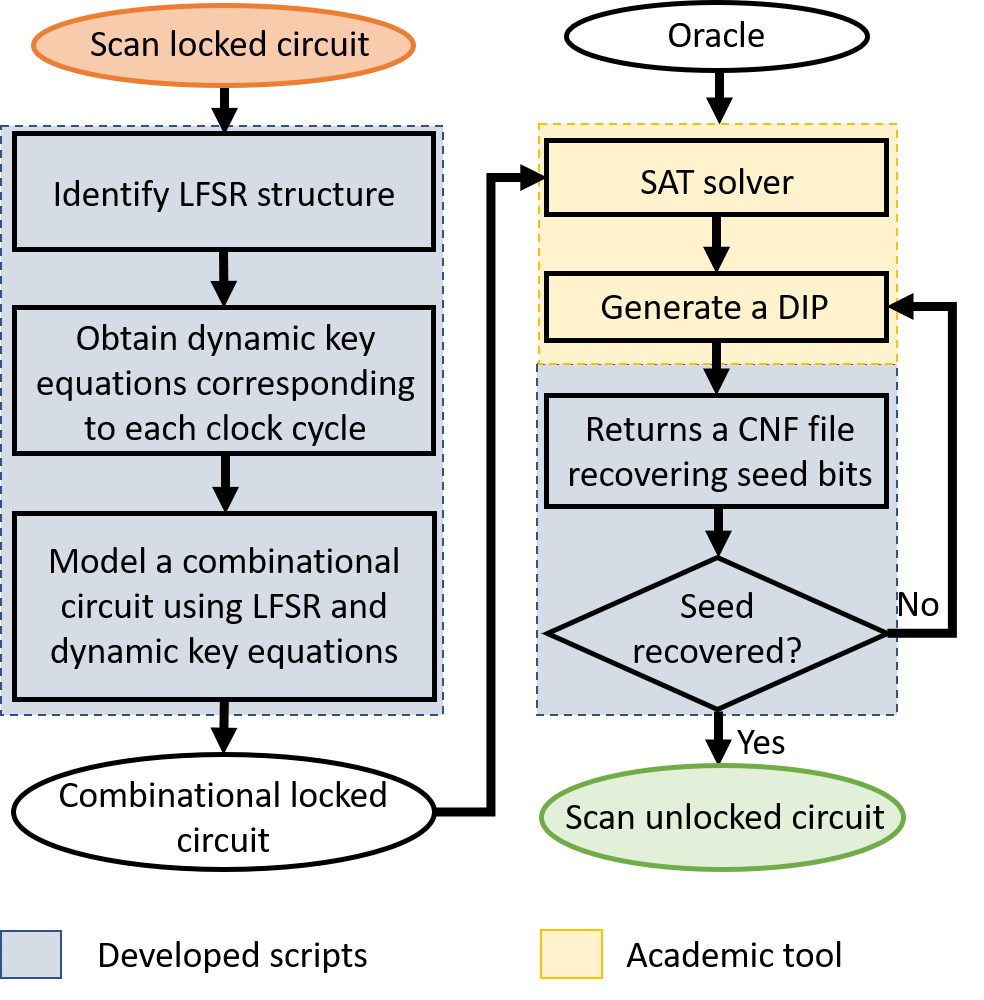}
    \caption{Flowchart for the proposed DynUnlock attack.}
    \label{fig:model}
\end{figure}

\begin{figure}
    \centering
    \includegraphics[width=0.45\textwidth]{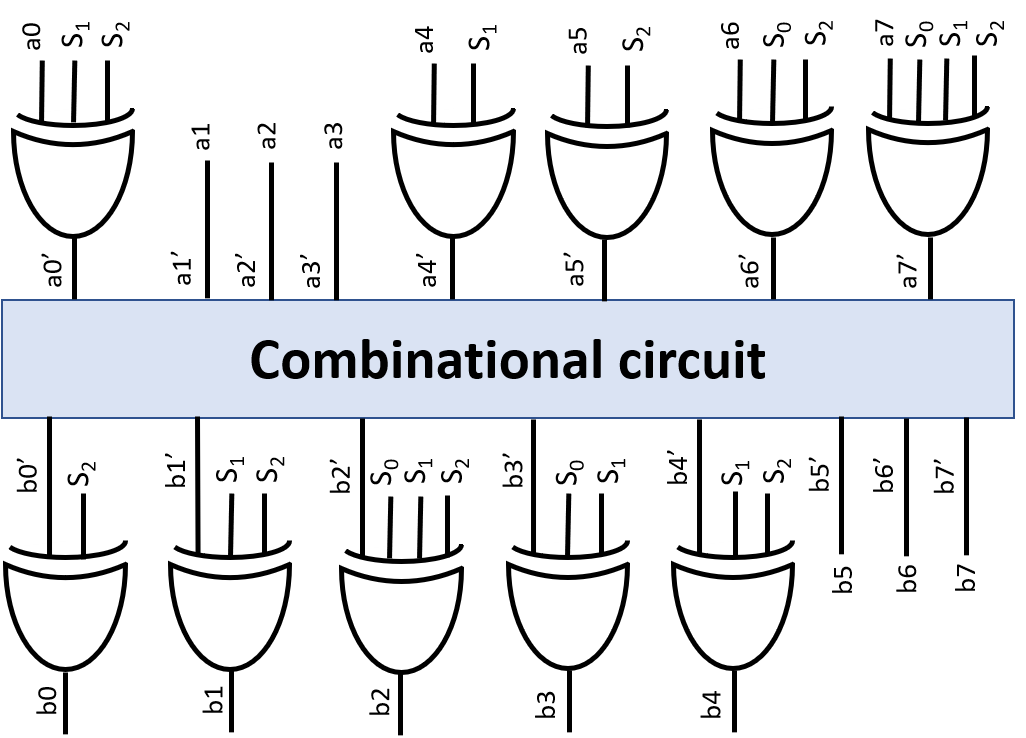}
    \caption{Combinational modeling of scan locked $s208$ circuit in Fig.~\ref{fig:eg} into a SAT attack compatible combinational circuit. $s_0$, $s_1$, and $s_2$ are the seed bits.}
    \label{fig:comb}
\end{figure}

\subsection{Attack Methodology}\label{sec:attack}
\textbf{Flowchart.}
As discussed before, we aim to recover the PRNG or LFSR seed and not the secret key ($SK$) for this particular case.\footnote{Here we assume that the PRNG is designed using an LFSR.} But, the methodology can be extended to other dynamic scan locking techniques whose seed is the secret key~\cite{DOS}.
We first start by reverse-engineering the LFSR circuit and obtaining the equations corresponding to each clock cycle. 
Next, we determine the location of key gates inserted between the scan flops and model this sequential logic circuit into a combinational circuit with scan flops replaced with inputs and outputs. As per the architecture, scan patterns are shifted in and out serially whereas key inputs are applied in parallel. 
Figure~\ref{fig:model} explains the attack methodology in a flowchart.

\begin{algorithm}[tb!]
% \footnotesize
\KwIn{$seed$, $a$, $b'$} 
\KwOut{$a'$, $b$} 
\textcolor{black}{
$FF$     = number of scan flops in the circuit\;
$seed$   = number of LFSR bits\;
$cycles$ = number of LFSR rounds\;                                                  %  \Comment{Additional one corresponds to the capture cycle}\\
$k^{y}$  = dynamic key for $y^{th}$ round\;}
% $FF_{loc}$ = location of locked flip-flops\;
$clk \gets 0$ \\                                                                     % \Comment{Forming equations 1-16} \\
\For{$i \gets 1$ to $cycles$}{
    $k^{i}_{0}$ $\gets$ $k^{i-1}_{1}$ $\oplus$ $k^{i-1}_{2}$ \\
    \For{$j \gets 1$ to $seed$} {
        $k^{i}_{j}$ $\gets$ $k^{i-1}_{j-1}$
    }
}
\For{$l \gets FF-1$ to $0$} {
    $i \gets 0$ \\
    $j \gets clk + 1$ \\
    \While{$i < seed$ and $j <= FF$}{
        $a_l \gets a_l \oplus k_{i}^{j}$\\
        $i \gets i + 1$\\
        $j \gets FF_{loc}[i]  + clk$ 
    }
    $a_l' \gets a_l$ \\
    $clk \gets clk + 1$
}
$clk \gets clk + 1$ \\
\For{$l \gets 0$ to $FF-1$} {
    $i \gets seed-1$ \\
    $j \gets FF_{loc}[i]  + clk$ \\
    \While{$i >= 0$ and $j > FF + 1$}{
        $b_l' \gets b_l' \oplus k_{i}^{j}$\\
        $i \gets i - 1$\\
        $j \gets FF_{loc}[i]  + clk$ 
    }
    $clk \gets clk - 1$\\
    $b_l \gets b_l'$ 
}
\caption{{\bf Combinational modeling of the dynamic scan locked circuit. $FF_{loc}[~]$ stores the positions of the scan locked flops. 
From Fig.~\ref{fig:eg}, $FF_{loc}[0:2] = (1, 2, 5)$.}
\label{alg:comb_model}}
\end{algorithm}

\textbf{Combinational Modeling.}
Consider scan locking on $s208$ ISCAS-89 benchmark shown in Fig.~\ref{fig:eg}, where the key gates are inserted after $1^{st}$, $2^{nd}$, and $5^{th}$ scan flops. The scan-in patterns are denoted by $a$ and scan-out patterns are denoted by $b$; $a'$ denotes the pattern delivered into an obfuscated scan chain and applied to the circuit, and $b'$ is the response of the circuit captured in the obfuscated scan chain. During shift operations where dynamic scan obfuscation is in place, $a$ turns into $a'$; and post-capture, $b'$ turns into $b$, both due to scan obfuscation.

Algorithm~\ref{alg:comb_model} shows the construction of LFSR keys from a seed, and a relationship between $a$ - $a'$ and $b$ - $b'$ in terms of the dynamic key bits. The first \texttt{for} loop corresponds to the LFSR equations, the second \texttt{for} loop corresponds to the relation between $a$ - $a'$, while the third \texttt{for} loop corresponds to the relation between $b$ - $b'$.
This completes our modeling step which results in a combinational locked circuit with seed bits acting as primary key inputs.

\begin{table}[!tb]
\caption{Results for scan locked circuits with 128-bit dynamic keys fed by an LFSR based on a secret seed. All ten benchmarks can be broken to obtain the complete 128-bit seed. For $s5378$, the attack recovers 16 seed candidates, whereas for $s13207$, it recovers 128 seed candidates, both of which can be easily brute forced to obtain the correct seed. }
\label{tab:results}
\centering
\setlength{\tabcolsep}{2pt}
\small\addtolength{\tabcolsep}{.0pt}
\footnotesize
\begin{tabular}{|c|c|c|c|c|c|}
\hline
\textbf{Benchmark} & \begin{tabular}[c]{@{}c@{}} \textbf{\# Scan flops}\footnote{The number of scan flops mentioned in the table are post-synthesis and may differ from the original number. This reduction is due to the fact that primary outputs were directly connected to flip flops without combinational cone between them.}\end{tabular}  & \begin{tabular}[c]{@{}c@{}}\textbf{\# Key}\\ \textbf{bits}\end{tabular} & \begin{tabular}[c]{@{}c@{}}\textbf{\# Seed}\\ \textbf{candidates}\end{tabular} & \textbf{\# Iterations} & \begin{tabular}[c]{@{}c@{}}\textbf{Execution}\\ \textbf{time (secs)}\end{tabular} \\ \hline
\hline
s5378      & 160            & 128         & 16                                                           & 17         & 41                                                                  \\ \hline
s13207     & 202            & 128         & 128                                                          & 4          & 27                                                                  \\ \hline
s15850     & 442            & 128         & 2                                                            & 4          & 89                                                                  \\ \hline
s38584     & 1,233           & 128         & 1                                                            & 3          & 219                                                                 \\ \hline
s38417     & 1,564           & 128         & 1                                                            & 7          & 342                                                                 \\ \hline
s35932     & 1,728           & 128         & 1                                                            & 1          & 254                                                                 \\ \hline
b20        & 429            & 128         & 1                                                            & 1          & 63                                                                  \\ \hline
b21        & 429            & 128         & 1                                                            & 1          & 54                                                                  \\ \hline
b22        & 611            & 128         & 1                                                            & 1          & 99                                                                  \\ \hline
b17        & 864            & 128         & 1                                                            & 1          & 86                                                                  \\ \hline
\end{tabular}
\end{table}

\textbf{SAT attack.}
Once modeling is complete, the combinational locked circuit is fed to a SAT tool~\cite{SAT} as shown in Fig.~\ref{fig:model}, which provides a DIP and its corresponding output pattern. We modify the code-base to dump a conjunctive normal form (CNF) after each iteration, which may reveal some of the seed bits. Unlike~\cite{ScanSAT_TETC}, we can carry out our attack for just one capture cycle. To recover more bits, we restart the LFSR circuit and obtain a new DIP and its corresponding output pattern from the SAT tool, and recover more seed bits. We repeat the restart step until all the seed bits have been recovered, or the remaining seed bits can be brute-forced. Even if the number of remaining seed bits is large for brute force, we can obtain a combinational locked circuit for a new capture cycle and carry over the seed information recovered from previous capture cycles to either recover the entire seed or reduce the brute force effort. We, however, have not come across any benchmark, locked with a key of a practical size, where a second capture was required, as can be seen in Section~\ref{sec:results}.
\section{Experimental Results}\label{sec:results}
\begin{table*}[!tb]
\caption{Results for $s38584$, $s38417$, and $s35932$ benchmarks locked using larger keys. We observe that even for a key size as large as 368 bits, there are only at most 16 seed candidates, which can be easily brute forced to recover the correct seed. The maximum time taken by any benchmark ($s38417$ for a key as large as 336 bits) is less than 23 hours. }
\label{tab:results_multiplekeys}
\centering
\setlength{\tabcolsep}{2pt}
\small\addtolength{\tabcolsep}{2pt}
\footnotesize
\begin{tabular}{|c|c|c|c|c|c|c|c|c|c|}
\hline
\multirow{2}{*}{\textbf{Key bits}} & \multicolumn{3}{c|}{\textbf{s38584}}                                                                                                                                       & \multicolumn{3}{c|}{\textbf{s38417}}                                                                                                                                      & \multicolumn{3}{c|}{\textbf{s35932}}                                                                                                                                       \\ \cline{2-10} 
                                   & \textbf{\begin{tabular}[c]{@{}c@{}}\# Seed \\ candidates\end{tabular}} & \textbf{\# Iterations} & \textbf{\begin{tabular}[c]{@{}c@{}}Execution\\ time (secs)\end{tabular}} & \textbf{\begin{tabular}[c]{@{}c@{}}\# Seed\\ candidates\end{tabular}} & \textbf{\# Iterations} & \textbf{\begin{tabular}[c]{@{}c@{}}Execution\\ time (secs)\end{tabular}} & \textbf{\begin{tabular}[c]{@{}c@{}}\# Seed \\ candidates\end{tabular}} & \textbf{\# Iterations} & \textbf{\begin{tabular}[c]{@{}c@{}}Execution\\ time (secs)\end{tabular}} \\ \hline
\textbf{144}                       & 1                                                                      & 3                      & 925                                                                      & 1                                                                     & 6                      & 862                                                                      & 1                                                                      & 1                      & 281                                                                      \\ \hline
\textbf{160}                       & 1                                                                      & 2                      & 557                                                                      & 1                                                                     & 5                      & 583                                                                      & 1                                                                      & 1                      & 634                                                                      \\ \hline
\textbf{176}                       & 1                                                                      & 2                      & 1,175                                                                     & 1                                                                     & 6                      & 1,711                                                                     & 1                                                                      & 1                      & 372                                                                      \\ \hline
\textbf{192}                       & 1                                                                      & 4                      & 872                                                                      & 1                                                                     & 6                      & 945                                                                      & 1                                                                      & 1                      & 618                                                                      \\ \hline
\textbf{208}                       & 1                                                                      & 5                      & 4,897                                                                     & 1                                                                     & 6                      & 1,947                                                                     & 1                                                                      & 1                      & 597                                                                      \\ \hline
\textbf{224}                       & 1                                                                      & 5                      & 4,792                                                                     & 1                                                                     & 4                      & 1,999                                                                     & 1                                                                      & 1                      & 1,007                                                                     \\ \hline
\textbf{240}                       & 1                                                                      & 6                      & 2,880                                                                     & 1                                                                     & 5                      & 2,252                                                                     & 1                                                                      & 1                      & 810                                                                      \\ \hline
\textbf{256}                       & 1                                                                      & 7                      & 9,219                                                                    & 2                                                                     & 7                      & 16,220                                                                & 1                                                                      & 1                      & 832                                                                      \\ \hline
\textbf{272}                       & 1                                                                      & 4                      & 2,831                                                                     & 4                                                                     & 7                      & 14,603                                                                   & 1                                                                      & 1                      & 1,364                                                                     \\ \hline
\textbf{288}                       & 1                                                                      & 7                      & 15,025                                                                  & 16                                                                      & 9                       &  24,546                                                                       & 1                                                                      & 1                      & 2,657                                                                     \\ \hline
\textbf{304}                       & 4                                                                      & 2                      & 6,465                                                                     & 16                                                                      & 14                       &  33,591                                                                       & 1                                                                      & 1                      & 1,881                                                                     \\ \hline
\textbf{320}                       & 4                                                                      & 6                      & 12,745                                                                   & 16                                                                    & 21                     & 62,135                                                                   & 1                                                                      & 1                      & 2,992                                                                     \\ \hline
\textbf{336}                       & 4                                                                      & 5                      & 10,678                                                                   & 16                                                                      & 17                       & 81,504                                                                         & 1                                                                      & 1                      & 2,008                                                                     \\ \hline
\textbf{352}                       & 4                                                                      & 5                      & 11,502                                                                   &  16                                                                     & 24                       & 74,140                                                                        & 1                                                                      & 1                      & 2,270                                                                     \\ \hline
\textbf{368}                       & 4                                                                      & 4                      & 11,173                                                                   & 16                                                                       &  27                      &   70,591                                                                      & 1                                                                      & 1                      & 3,231                                                                     \\ \hline
\end{tabular}
\end{table*}

\textbf{Setup.}
We conducted our DynUnlock attack on six ISCAS-89~\cite{iscas89} and four ITC-99~\cite{itc99} benchmarks with dynamically locked scan chains with a 128-bit key. 
Synopsys Design Compiler was used to synthesize these benchmarks.
\textcolor{black}{All the experiments have been carried out on a $24$-core Intel Xeon processor running at $3.33$ GHz with $96$ GB RAM. }

\begin{comment}
\begin{table}[!tb]
\caption{ISCAS-89 and ITC-99 benchmark statistics. The benchmarks are first arranged suite-wise and then based on the number of scan flops in the circuit.}
\label{tab:benchmarks}
\centering
\setlength{\tabcolsep}{2pt}
\small\addtolength{\tabcolsep}{.0pt}
\footnotesize
\begin{tabular}{|c|c|c|c|c|}
\hline
\textbf{Benchmark} & \textbf{\# Gates} & \textbf{\# Inputs} & \textbf{\# Outputs} & \textbf{\# Scan flops} \\ \hline
\hline
s5378      & 2,779     & 35        & 49         & 179           \\ \hline
s13207     & 7,951     & 62        & 152        & 638           \\ \hline
s15850     & 9,772     & 77        & 150        & 534           \\ \hline
s38584     & 19,253    & 38        & 304        & 1,426          \\ \hline
s38417     & 22,179    & 28        & 106        & 1,636          \\ \hline
s35932     & 16,065    & 35        & 320        & 1,728          \\ \hline
b20        & 17,158    & 32        & 22         & 490           \\ \hline
b21        & 17,482    & 32        & 22         & 490           \\ \hline
b22        & 25,460    & 32        & 22         & 735           \\ \hline
b17        & 27,852    & 37        & 97         & 1,415          \\ \hline
\end{tabular}
\end{table}
\end{comment}
\textbf{Attack results.}
Our attack results are given in Table~\ref{tab:results} which shows the number of seed candidates recovered, number of iterations required by the lingeling SAT solver, and the attack execution time. As we can see, DynUnlock successfully unlocks all the circuits by recovering the 128-bit seed. For benchmarks $s5378$, $s13207$, and $s15850$, the attack obtains more than one seed candidate, which highlights that with just one capture cycle, SAT tool was not able to resolve all the CNF equations. However, these seed candidates are very few in number which can be easily brute forced to obtain the correct seed. For the rest of the benchmarks, we directly obtain the correct seed. All these benchmarks are run for 10 different LFSR seeds, and the number of seed candidates, number of iterations, and the execution times are averaged over these 10 runs.

\textbf{Attack scalability.}
In our attack, we include the LFSR structure in the combinational modeling. All the seed bits are correlated with other seed bits as shown in Fig.~\ref{fig:comb}. Hence,  the SAT attack sometimes resolves only these clauses and leaves with multiple values for the variables in these clauses, thereby, increasing the seed candidates.
For larger circuits, we obtain only one seed candidate. Intuitively, in a larger circuit with a larger number of scan flops, attack success should be higher as the seed bits will repeat for a larger number of times. We confirm this intuition from the results in Table~\ref{tab:results}. Furthermore, for larger circuits ($s38584$, $s38417$, and $s35932$), even if the key-size is 240 bits, we still recover one seed as shown in Table~\ref{tab:results_multiplekeys}, and upto key sizes of 368, we obtain at most 16 seed candidates, which can be easily brute forced to recover the correct seed. \textbf{Thus, our attack is scalable with number of scan flops as well as with increasing key sizes.} 
\section{Discussion and Conclusion}
Among the scan locking solutions that follow the same threat model used in logic locking, the only defenses that our attack cannot circumvent are those that incorporate cryptographic functions~\cite{stream_vs_block, da2018preventing} or PUF structures~\cite{Bias_PUF} to generate dynamic keys. Our attack cannot model such modules into their combinational logic equivalent. While this is a limitation of our attack, we note that cryptographic functions incur significant area and power cost, limiting the use of such defenses to chips that have these blocks already. 

In this paper, we present a novel attack which circumvents the most dynamic case of scan locking proposed in~\cite{ImprovedEFF}. We model the sequential circuit into a locked combinational circuit on which the SAT attack~\cite{SAT} can be applied. We conducted experiments on large circuits in ISCAS-89 and ITC-99 benchmark suites and \textbf{recovered the LFSR seed within seven minutes for all the benchmarks under consideration.} We also evaluated the scalability of our attack with increasing key sizes and \textbf{recovered the seed using at most 27 iterations and within 23 hours for all the 15 cases for key sizes as large as 368 bits.}
Our attack can break any version of dynamic scan locking. With our attack process, we will never run out of iterations, as the attack will always provide seed candidates, if not the correct unique seed. 

\section*{Acknowledgement}
The authors would like to acknowledge Satwik Patnaik for his valuable feedback. This work was supported by New York University NY/Abu Dhabi Center for Cyber Security and Intel Corporation.

\bibliographystyle{IEEEtran}
\footnotesize{
    \bibliography{main.bib}
}

\end{document}